# A No-Go Theorem for Matter-wave Interferometry with Application to the Neutron's Electric-dipole Moment

Murray Peshkin*

*Physics Division, Argonne National Laboratory, Argonne, IL 60439-4843 USA*

## Abstract

A theorem that relies only on the unitary property of the Schroedinger equation and not upon any classical or semi-classical approximation negates some, but not all, suggestions that have been made for measuring the neutron's electric-dipole moment by interferometry.

PACS: 03.75.-g

*E-mail: peshkin@anl.gov

# I. MOTIVATION

It is now known that the neutron's electric dipole moment (EDM) $\mu_e$ is indistinguishable from zero with experimental sensitivity around $10^{-25}$ $e$-cm [1], and it is anticipated that the sensitivity of near-future experiments based on the Ramsey method cannot be improved by more than one or two orders of magnitude [2]. That in-principle limitation is reviewed in Appendix A. It applies to any experiment that deduces the EDM from an energy shift, or equivalently from a precession frequency, in an applied electric field. Mainly for that reason, interferometric experiments that appear to have at least the possibility of achieving much greater sensitivity have been proposed. However, calculations of the desired interference effects, all of them based on semi-classical approximations that are good enough for ordinary interferometry but not necessarily good enough for the delicate situations involved in the proposed EDM measurements, have not all arrived at the same conclusions. Here I give a simple no-go theorem, based on the unitarity of the Schroedinger equation and not on any semi-classical approximation, which rules out some interferometry measurements of the neutron's EDM but not others.

# II. PHASE-SHIFT GROWTH IN A DRIFT SPACE

Consider two very different interferometry schemes. In the Colella-Overhauser-Werner (COW) experiment [3], a neutron enters a Mach-Zehnder interferometer in which the two arms are at different gravitational potentials, one above the other. The two partial wave packets in the two arms are centered around slightly different wave numbers. The packets acquire a relative-phase shift equal to $(\Delta k)L$, where $\Delta k$ is the difference between the central wave numbers of the two packets and $L$ is the length of the drift space. The two packets then strike the two sides of a half-reflecting mirror and the intensities of the two emerging beams depend upon the phase shift, which is proportional to the drift length $L$.

The second scheme is illustrated by one version of a proposal of Freedman, Ringo, and Dombeck (FRD) for measuring the neutron's EDM [4]. In that proposal, illustrated schematically in Fig. 1, ultracold neutrons are admitted to an accelerator box containing a strong electric field gradient that accelerates neutrons with $\sigma_x = +1$ in the $+x$ direction and those with $\sigma_x = -1$ in the $-x$ direction, both proportionately to $\mu_e$. Exactly how that is to be done is immaterial for present purposes. A neutron admitted at time $t_0$ with its spin polarized in the $+y$ direction is represented by a wave function

$$\Psi(t_0) = \psi_y(r, t_0)\chi_{y+} = \frac{1}{\sqrt{2}}\left[\psi_+(r, t_0)\chi_+ + \psi_-(r, t_0)\chi_-\right], \tag{2.1}$$

where $\psi_y(r, t_0)$ is some wave packet, $\chi_{y+}$ is the spin function for $\sigma_{y+} = +1$, and $\chi_\pm$ are the spin functions for $\sigma_x = \pm 1$, rsp. The two partial wave packets $\psi_+$ and $\psi_-$ are both equal to $\psi_y$ at time $t_0$, but later they will be slightly different from each other, their centroids separating in the $x$ direction by an amount proportional to $\mu_e$ and to the square of the time spent in the accelerator box. That separation is governed by the Hamiltonian

$$H = \frac{\boldsymbol{p}^2}{2m} + mgz - \mu_m\,\boldsymbol{\sigma}\cdot\boldsymbol{B} - \mu_e(\boldsymbol{\sigma}\cdot\boldsymbol{\nabla})\boldsymbol{E}, \tag{2.2}$$

which results in

$$\Psi(t) = \frac{1}{\sqrt{2}}\left[e^{+i\omega_g(t-t_0)/2}\psi_+(r,t)\chi_+ + e^{-i\omega_g(t-t_0)/2}\psi_-(r,t)\chi_-\right], \tag{2.3}$$

where $\omega_g$ is the neutron's spin precession frequency in a weak uniform magnetic guide field $B$ in the $z$ direction that pervades the entire experiment for practical reasons. The spatial wave functions $\psi_\pm$ are governed by the Hamiltonian (2.2) except that the magnetic field term must be removed.

FRD estimate that under useful assumptions the two wave packets will separate by femtometers in the accelerator box, much less than the width of the packets, and the goal of the proposed experiment is first to amplify that separation and then to measure it by interferometry. The amplification is provided by removing the lid of the accelerator box after some time so that the neutrons come out the top of the box and strike a 45-degree mirror, then bounce along through a horizontal drift tube. Ideally, after leaving the box the neutrons experience no electric field gradient. The initially horizontal separation of the two partial wave packets is converted to a partly vertical separation by reflection from the mirror. Representative classical orbits are shown in Fig. 1. The average potential energy separation between the two packets, while still proportional to $\mu_e$, is now also proportional to the weight of the neutron and orders of magnitude greater than $\mu_e E$ for any achievable electric field. FRD then calculate a growing phase shift, proportional to the length of the drift tube between the two wave packets. That calculation is based on line integrals $\int \boldsymbol{k} \cdot d\boldsymbol{r}$ of the semi-classical wave number along the classical orbit and it is not repeated here. The two partial waves acquire a growing phase difference as they progress down the drift space because the orbit that bounces lower has greater average wave number. The so-calculated phase shift is proportional to the EDM and to the length of the drift space. Others have obtained a different result from the semi-classical approximation and found no phase shift enhancement as the neutron progresses down the drift space [5].

In fact, the relevant phase shift depends upon what measurement is made at the end of the drift space. In the simplest version of FRD, the neutron is caught in a box at the end of the drift space and a component of its polarization $\boldsymbol{P}$ is measured.

$$\boldsymbol{P}(t) = \int \Psi(r,t)^* \boldsymbol{\sigma} \Psi(r,t) d^3r \tag{2.4}$$

The component of $\boldsymbol{P}$ in a direction in the $yz$ plane at some angle $\vartheta$ to the $y$ axis is given by

$$P_\vartheta(t) = Re\left\{e^{-i\vartheta}e^{-i\omega_g(t-t_0)}I(t)\right\}, \tag{2.5}$$

where

$$I(t) = \int \psi_-(r,t)^* \psi_+(r,t) d^3r. \tag{2.6}$$

The factor $e^{-i\omega_g(t-t_0)}$ represents the precession in the magnetic guide field and the phase of $I(t)$ includes any additional precession.

## III. THE NO-GO THEOREM

The no-go theorem states that $I(t)$ is a constant of the motion in the absence of fields other than gravity and the magnetic guide field $B$. For the specific version of the FRD proposal described above, that says there is no precession proportional to $\mu_e$ after the neutron leaves the accelerator box, and no enhancement of the phase shift proportional to the length of the drift space.

*Proof* :

The Hamiltonian for the neutron's motion in the absence of fields except for gravity and the magnetic guide field $B$ is

$$H = \frac{\boldsymbol{p}^2}{2m} + mgz - \mu_m \, \boldsymbol{\sigma} \cdot \boldsymbol{B} \,. \tag{3.1}$$

Applying that to Eq. (2.3) gives

$$i\hbar \frac{\partial \psi_\pm}{\partial t} = \left[\frac{\boldsymbol{p}^2}{2m} + mgz\right]\psi_\pm \,. \tag{3.2}$$

Since $\psi_+$ and $\psi_-$ obey the same Schroedinger equation, the unitarity principle implies that $I(t)$ is a constant of the motion.

## IV. APPLICABILITY

The no-go theorem applies when the quantity measured is proportional to $I(t)$ and the two partial waves obey the same Schroedinger equation, as in the version of FRD described above. It also applies to some spin-flip detection schemes.

The COW experiment has formally a similar constant of the motion, with only the difference that there the spin can be ignored and one has just the two partial waves in the two arms of the interferometer. In that case, $I(t)$ vanishes and it remains equal to zero when the wave packets have been recombined and are emerging from the interferometer. However, the no-go theorem does not speak usefully to that experiment because there the phase shift shows up as an overlap integral between two partial wave packets in one emergent beam only, not as the conserved integral over all space. For the same reason, the theorem also does not speak usefully to versions of the FRD proposal in which the phase shift is measured with a Mach-Zehnder interferometer instead of a polarimeter [6]. The physical issue in both cases is that a half-reflecting mirror is inserted between the two beams when they are recombined, and that measures a phase shift different from $I(t)$.

A suggestion of Anandan also avoids the no-go theorem by using a Mach-Zehnder interferometer [7]. There, one partial wave is subjected to an electric field in the direction of the neutron's polarization. That does give a phase shift between the two beams, measurable in principle when they are reunited. The no-go theorem does not speak to that arrangement because the neutrons are continually in an-external electric field that causes the Hamiltonian to depend upon the neutron's spin. However, the uncertainty-principle limit for the Anandan arrangement is the same as that for the Ramsey method. (See Appendix B.)

Dombeck [8] has recently proposed another interferometry measurement of the neutron's EDM to which the no-go theorem does not speak because the experiment does not rely on an electric-field-free drift space. That proposal does appear to beat the uncertainty-principle limit of the Ramsey method.

Finally, one can imagine a variant of FRD in which the polarization measurement is made in a small range of $x$ near some $x_0$, possibly even as a function of time, instead of at all $x$ at one time, to avoid the integral over all space in $I(t)$. That avoids the no-go theorem in principle, but in practice that variant fails by orders of magnitude to be useful for measuring the neutron's EDM.

## ACKNOWLEDGMENTS

I thank G. R. Ringo and T. W. Dombeck for very helpful discussions about this work. This work is supported by the U. S. Department of Energy, Nuclear Physics Division, under contract W-31-109-ENG-38.

FIGURES

FIG. 1. The FRD proposal. The dashed line on top of the accelerator box represents its removable lid. Representative orbits of the centers of the two partial wave packets originate from cirles representing the wave packets. The separation of the two packets is vastly exaggerated in comparison with their size.

## APPENDIX A: The "Uncertainty Principle" Limit

For completeness, I repeat here briefly the explanation of what has come to be called the uncertainty principle limit on the sensitivity of the Ramsey method [2]. A neutron's spin, initially pointing in the $x$ direction, precesses in the $xy$ plane under the influence of an applied magnetic field in the $z$ direction. An added electric field of strength $E$ in the $z$ direction causes an incremental rate of precession $\omega$ given by

$$\omega = \frac{2\mu_e E}{\hbar} \tag{A.1}$$

and an incremental integrated precession angle $\phi$ given by

$$\phi = \omega T = \frac{2\mu_e E}{\hbar} T, \tag{A.2}$$

$T$ being the time of exposure to the electric field.

For a single neutron the statistical error in $\phi$ is around one radian, so the statistical error of the EDM measurement is about $\approx \hbar/ET$. Then for N neutrons,

$$\delta\mu_e \approx \frac{\hbar}{ET\sqrt{N}}. \tag{A.3}$$

In the practical case $T$ is limited by the 15-minute mean life of the neutron, $E$ is limited to around 50,000 V/cm, and $N$ can be perhaps $10^8$. Using those numbers

$$\delta\mu_e \approx 10^{-27} e - cm \tag{A.4}$$

appears to be the limiting sensitivity, not considering systematic errors. The experimenters are not optimistic about the possibility of increasing $E$ or $N$ enough to make a major improvement in the limiting $\delta\mu_e$.

## APPENDIX B: The Anandan Interferometer

In the Anandan arrangement one arm of a Mach-Zehnder interferometer contains a region of length $L$ where there is a stationary electric field $E$ in the direction of the neutron's polarization. The neutron is accelerated by the electric field gradient as it enters and leaves the field region. There is no change in energy, but the neutron's momentum changes by

$$\Delta p = \frac{m}{p} \mu_e E. \tag{B.1}$$

That gives a phase shift

$$\Phi = \frac{\Delta p}{\hbar} L = \frac{mL}{\hbar p} \mu_e E. \tag{B.2}$$

As in the Ramsey method, the statistical error $\delta\Phi$ is around one radian divided by the square root of the number of neutrons. The length of the interferometer cannot be much greater than the speed of the neutron times its mean life $T$. Then the in-principle limiting sensitivity is given by

$$\delta\mu_e \approx \frac{\hbar}{ET\sqrt{N}}, \tag{B.3}$$

the same as in the Ramsey method.

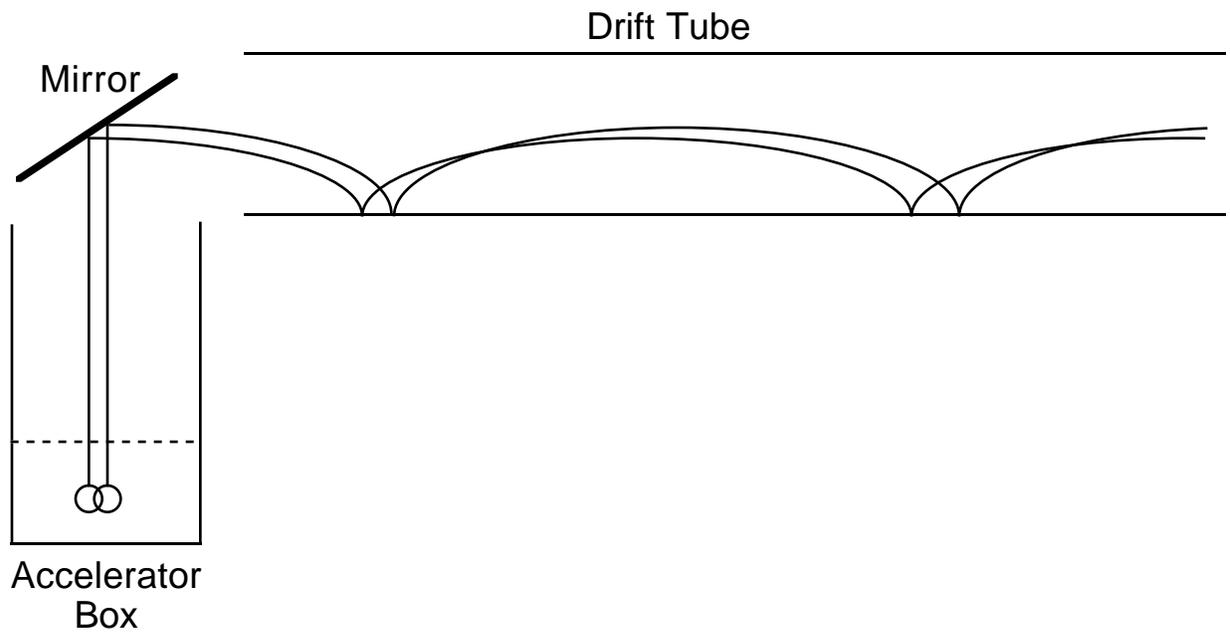